\documentclass[aps, twocolumn, showpacs]{revtex4-1}
\usepackage{amsmath}
\begin{document}
\title{Coulomb- quantum oscillator correspondence in two dimension, pure gauge field and half-quantized vortex}
\author{S. C. Tiwari \\
Department of Physics, Institute of Science, Banaras Hindu University, Varanasi 221005, and \\ Institute of Natural Philosophy \\
Varanasi India\\}
\begin{abstract}
Isotropic oscillator and Coulomb problems are known to have interesting correspondence. We focus on 2D quantum problems
and present complete treatment on the correspondence including the Schroedinger equation, eigenfunctions and eigenvalues,
and the integrals of motion. We find only partial equivalence. The wavefunction correspondence is examined introducing
local gauge transformation and the emergence of half-quantized vortex with the associated spin-half is established. Vortex
structure of the electron proposed by us and the origin of charge are discussed in the present context. Outlook on
the implications for QCD and hadron spectrum is outlined.
 
\end{abstract}
\maketitle

\section{\bf Introduction}

Spatio-temporal bounded structures and topological obstructions in the physical space-time as the fundamental constituents of the 
elementary particles and their interactions have been explored by the author in the past for developing an alternative to the Standard Model;
see recent paper \cite{1} and the references cited therein. The idea that the electron charge has origin in angular momentum, and that space-time
vortex in 2+1 dimension is the basic elementary topological object have been the key elements in this approach. The occurrence of SU(2) group 
in 2D isotropic oscillator noted by Jauch and Hill in 1940 \cite{2} may have deep physical significance \cite{3}. The proposition of the existence of
vortex in 2D oscillator \cite{4} and the analysis of the connection between the Coulomb (or Kepler) and the isotropic oscillator problems
throws new light on the meaning of spin and coupling constant. We emphasize that prior to Pauli and Dirac spinors in physics Elie
Cartan in 1913 had given geometric origin of spinors. In recent years a vast literature on classical and 
geometric interpretations of spin and its origin has appeared, for some of the references see \cite{1}. However the scope of the present paper is restriced
to the symmetries in 2D oscillator.

Bergmann and Frishman \cite{5} established a relation between the hydrogen atom and isotropic oscillator in even dimensions from 2 to 4l+4, where
l is orbital angular momentum quantum number. McIntosh \cite{6} mentions that Schwinger in an unpublished work had earlier obtained a connection
between the Coulomb and the oscillator radial equations; in fact, the radial coordinate transformation used in \cite{5} is same as that of Schwinger. 
A general mapping for d-dimensional Coulomb problem and D-dimensional isotropic oscillator problem, and its supersymmetric extension have also
been discussed \cite{7}. There exists a vast literature on this subject, however we focus on d=D=2 dimensional nonrelativistic theory.

Symmetry and degeneracy based on the integrals of motion \cite{2, 6} and the Coulomb-oscillator connection based on the radial Schroedinger
equation following the approach of \cite{5}  are usually discussed in the literature. Though Ter-Antonyan \cite{8} does not consider 
the integrals of motion, the Levi-Civita transformation used by him for 2D problem includes both radial coordinate and polar angle. Author terms
his approach nonstandard since the energy eigenvalue of the quantum oscillator is assumed fixed and the frequency is quantized, however this approach 
is quite insightful. In an interesting paper \cite{9} one of the integrals of motion, namely, the angular momentum under the 
Levi-Civita transformation is shown to throw light on the half-integral spin. A full discussion of 2D problem that includes integrals of motion,
stationary state wavefunction, Schroedinger equation and energy eigenvalues does not exist in the literature. Incorporating the role of pure
gauge field and vortex \cite{4} we present a complete treatment of the 2D Coulomb-oscillator correspondence in this paper. One of the important 
results obtained from this analysis is the concrete realization of spin. In addition a new insight is gained on the nature of the coupling constant.

For the self-contained discussion we explain topological ideas for vortex, specially in 2D problem in the next section. In Section 3 the 
full treatment of the 2D Coulomb-oscillator correspondence is presented. It is shown that the equivalence is only partial. The role of pure
gauge field in the generation of half-quantized vortex is investigated in Section 4. Physical interpretation and new implications for QCD and 
hadron spectrum constitute the last section. 

\section{\bf Vortex and Topology}

The idea that vortex appears in 2D isotropic quantum oscillator \cite{4} is important, however its foundation could be elaborated more generally. 
Let us recall that the term 'vortex' in optics and quantum theory has origin in the fluid dynamics. A singular vortex in fluid dynamics has 
a simplified representation in a 2D potential theory such that the negative gradient of the potential determines the velocity field. Assuming
potential function proportional to the polar angle in polar coordinate system $(r, \phi)$, the azimuthal velocity field is proportional to 
$\frac{1}{r}$, where $r\neq 0$. Geometrically one has a punctured plane $R^2 -\{0\}$, and it is topologically equivalent to a cylinder. In contrast 
to the cartesian coordinates $(x, y)$ the description in polar coordinates has nontrivial topological aspects: the angle $\phi$ has to be restricted 
to $[0, 2 \pi)$ since arctangent is a multi-valued function. In this case the potential function is not globally defined, i. e. encircling the origin
it changes by integral multiple of $2 \pi$, as a result the velocity field in the language of exterior differential forms is closed but not exact.
For an intuitive physical understanding of topology in physics we refer to \cite{10, 11, 12}.

In quantum theory Madelung's hydrodynamical interpretation of the Schroedinger equation provides a natural concept of quantized vortex.
Note that the role of nodes at which the wavefunction becomes zero is important, however the wavefunction must be complex. For example,
quantum oscillator in 1D has nodes, but it is real and there are no vortices. Hydrogen atom wavefunction $\Psi_{nlm}(r,\theta,\phi)$ vanishes 
at the origin $r=0$ except for the s-states, i. e. $l=0$. Nodal line is along z-axis around which multi-valued phase gives rise to quantized 
phase vortex \cite{13}. The argument that the vortices exist for a complex wavefunction possessing zeros is straightforward utilizing the definition of
probability current density, and its conservation law. Consider Schroedinger wavefunction
\begin{equation}
\Psi({\bf r},t) = \sqrt{\rho } e^{\frac{i S}{\hbar}}
\end{equation}
then the probability current density defined by
\begin{equation}
{\bf J} = \frac{i \hbar}{2 m} [\Psi {\bf \nabla} \Psi^* - \Psi^* {\bf \nabla} \Psi] 
\end{equation}
satisfies the continuity equation
\begin{equation}
{\bf \nabla}.{\bf J} + \frac{\partial \rho}{\partial t} =0 
\end{equation}
In analogy to hydrodynamics the velocity field is defined to be
\begin{equation}
{\bf v} =\frac{{\bf J}}{\rho} = \frac{{\bf \nabla} S}{m} 
\end{equation}
Now $\Psi$ is a single-valued function, it follows that for multi-valued phase function $S$ there exist quantized vortices.

In 2D oscillator the stationary state wavefunction, i. e. energy eigenfunction solving the Schroedinger equation in cartesian coordinates is just the 
product of 1D oscillator eigenfunctions; the unnormalized wavefunction is
\begin{equation}
\Psi_{n,m} = H_n(\xi x) H_m(\xi y) e^{- \xi^2 (x^2+y^2)/2} 
\end{equation}
where $\xi^2 =\frac{m \omega}{\hbar}$, and $H_n$ is a Hermite polynomial of order n. This solution has two interesting features: 
1) it is not eigenfunction of the angular momentum operator
\begin{equation}
L_z =-i\hbar (x \frac{\partial}{\partial y} - y \frac{\partial}{\partial x})
\end{equation}
except for the ground state, i. e. trivially $l=0$, and 2) it is real, therefore the velocity field defined by (4) vanishes, and
there is no vortex though the wavefunction has nodes. The same problem solved in polar coordinates results into energy eigenfunction
with well-defined angular momentum. The phase factor $e^{i l \phi}$ gives rise to the function $\frac{S}{\hbar} = l \phi$ responsible for
the quantized circulation of the vortex
\begin{equation}
 \oint {\bf v}. d{\bf r} = 2 \pi N \frac{\hbar}{m} 
\end{equation}
Here $N= 0, \pm 1, \pm 2 ....$.

The choice of the coordinate system seems to alter the
physical nature of the wavefunction. Formally one could
use the degeneracy and the superposition principle of
quantum mechanics to construct $L_z$ eigenstates combin-
ing different degenerate $\Psi_{n,m}$ states in a suitable manner.
We have argued that there must exist a physical mech-
anism to transform vortex-free zero angular momentum
states to nonzero angular momentum vortex states \cite{4}.

\section{\bf Coulomb-Oscillator correspondence}

The transformation of the Coulomb radial equation to the oscillator radial equation in quantum theory affected by radial coordinate
transformation \cite{5, 7} is a remarkable result. Ter-Antonyan \cite{8} suggests dyon-oscillator duality analogy with Seiberg-Witten
duality re-visiting the Coulomb-oscillator correspondence. Fu et al \cite{9} discuss 2D systems applying the Levi-Civita coordinate
transformation \cite{8} to analyze relativistic Dirac and Klein-Gordon equations. Here we  consider only 2D nonrelativistic Schroedinger
equation for the Coulomb and isotropic oscillator problems.

The Hamiltonian of the isotropic oscillator
\begin{equation}
H^o = \frac{1}{2 m} ( p_u^2 + p_v^2) + \frac{1}{2} m \omega^2 (u^2 +  v^2)
\end{equation}
leads to the Schroedinger equation for the stationary states
\begin{equation}
H^o \Psi^o = E^o \Psi^o 
\end{equation}
There exist four constants of the motion \cite{2}; not all of them are independent as their maximum number could only be three. Besides
the Hamiltonian following three integrals \cite{2} are written here
\begin{equation}
 F^o_1 = -\frac{1}{2} (u p_v - v p_u) 
\end{equation}
\begin{equation}
F^o_2 = -\frac{1}{4} [ (m \omega)^{-1} (p_u^2 - p_v^2) + m \omega (u^2 - v^2)] 
\end{equation}
\begin{equation}
F^o_3 = -\frac{1}{2} [ (m \omega)^{-1} p_u p_v +  m \omega u v ] 
\end{equation}
The integrals commute with $H^o$ and satisfy the following commutation rules
\begin{equation}
[F^o_i , F^o_j ] = i \hbar \epsilon_{ijk} F^o_k 
\end{equation}
The Schroedinger equation (9) written in explicit form 
\begin{equation}
[-\frac{\hbar^2}{2 m} (\frac{\partial^2 }{\partial u^2} + \frac{\partial^2}{\partial v^2}) + \frac{1}{2} m \omega^2 (u^2 +v^2)] \Psi^o = E^o \Psi^o
\end{equation}
can be solved to obtain the energy eigenvalues
\begin{equation}
E^o_N = (n+m+1) \hbar \omega 
\end{equation}
along with the eigenfunctions having the form (5) consisting of the standard oscillator functions. The degree of degeneracy is $N+1$, where $N= n+m$.
Since N can take all positive integral values, the half-integral representation of the rotation group is allowed. Note that (13) is a closed Lie
algebra of the 3D rotation group but having a factor of half in (10)-(12).

We are interested in giving full treatment of the Coulomb-oscillator correspondence and need integrals of motion for the Coulomb problem.
Jauch and Hill \cite{2}  discuss 2D Coulomb problem as a separate example. Let us introduce the Levi-Civita transformation \cite{8,9}
defined by
\begin{equation}
x = u^2 - v^2 
\end{equation}
\begin{equation}
y = 2 u v 
\end{equation}
\begin{equation}
r = (x^2+y^2)^{\frac{1}{2}} = u^2 +v^2 = R^2 
\end{equation}
The Hamiltonian $H^o$ given by expression (8) transforms to 
\begin{equation}
\tilde{H}^o = \frac{4r}{2m} (p_x^2 + p_y^2) + \frac{1}{2} m \omega^2 r 
\end{equation}
On the other hand the Coulomb Hamiltonian reads
\begin{equation}
H^c = \frac{1}{2m} (p_x^2 + p_y^2) - \frac{e^2}{r} 
\end{equation}
having no connection with $\tilde{H}^o$, i. e. the expression (19). Interestingly the Schroedinger equation (14) under the 
transformation (16)-(18) can be arranged in the following form resembling the Coulomb Schroedinger equation
\begin{equation}
[-\frac{\hbar^2}{2\mu} (\frac{\partial^2}{\partial x^2} + \frac{\partial^2}{\partial y^2}) - \frac{\alpha}{r}] \Psi^c = E^c \Psi^c 
\end{equation}
\begin{equation}
 E^o = \alpha
\end{equation}
\begin{equation}
E^c = - 2 \mu \omega^2 
\end{equation}
where the coupling constant denoted by $\alpha$ depends on the energy, $m= 4 \mu$, and the negative sign in (23) signifies bound state Coulomb
problem.

Let us now consider orbital angular momentum (10) under the Levi-Civita transformation. We find
\begin{equation}
F^o_1 ~ \rightarrow ~ F^c_1 = (x p_y - y p_x) 
\end{equation}
Note the presence of the factor of half in the definition (10); Fu et al \cite{9} remark that this explains as to why
the half-integral values are allowed in the oscillator.

The nature of the remaining two integrals of motion (11) and (12) is quite delicate: apparently $F^o_2$ is related with energy difference
of the two 1D oscillators, while $F^o_3$ according to \cite{2} has 'no obvious physical significance'. McIntosh \cite{6} offers illuminating
interpretation in terms of the generators of the infinitesimal contact transformations such that $F^o_3$ generates infinitesimal changes 
in the eccentricity of the elliptic orbit keeping the sum of the semi-axes fixed, and $F^o_2$ affects the phase changes between the 
two 1D oscillators. In the orbit picture one may visualize $F^o_3$ changing a nearly circular orbit to a straight line and then to an ellipse
with an opposite sense after a rotation of $2\pi$. To return to the original orbit one needs $4\pi$ rotation, and this explains  two-valued
representation for the oscillator.

In the Coulomb problem too, we have elliptical orbits, however unlike the oscillator where the force center is at the center,
the force acts at a focus of the ellipse in this case. The integrals of motion \cite{2} include the ones constructed from the 
Laplace-Runge-Lenz-Pauli vector \cite{14}
\begin{equation}
F^c_2 = - \frac{1}{2 \mu e^2} (F^c_1 p_y +p_y F^c_1) +\frac{x}{r} 
\end{equation}
\begin{equation}
F^c_3 =  \frac{1}{2 \mu e^2} (F^c_1 p_x +p_x F^c_1) +\frac{y}{r}
\end{equation}
The Levi-Civita transformed expressions (11) and (12) are calculated to be
\begin{equation}
\tilde{F}^o_2 = - \frac{1}{4} [ (m\omega)^{-1} ( 4 x p_x^2 -4x p_y^2 + 8 y p_x p_y) + m\omega x 
\end{equation}
\begin{equation}
\tilde{F}^o_3 = - \frac{1}{2} [ (m\omega)^{-1} ( -2yp_x^2 + 2 y p_y^2 +4 x p_x p_y) + \frac{1}{2} m \omega y]
\end{equation}
Expressions (27) and (28) differ markedly from the Coulomb integrals of motion (25) and (26).

Disagreement on the equivalence of these integrals of motion may be understood from qualitative physical arguments.
First reason seems to be the displaced force center in the Coulomb problem, i. e. at the focus, and secondly the integrals of motion
$F^c_1, F^c_2, F^c_3$ do not form a closed algebra unlike that for the oscillator case. Since the Hamiltonian $H^c$ appears in the commutators
one assumes a sub-Hilbert space corresponding to a constant energy eigen value (which is less than zero for bound states) and arrives
at a closed Lie algebra for 3D rotation group SO(3). Note that in the case of 2D oscillator the integrals of motion generate SU(2) group \cite{2,6}.

Thus the full treatment carried out for the Coulomb-oscillator correspondence shows that there exists only a partial equivalence between them.
It may be asked if this result, in some sense, reflects the limitation of the Levi-Civita coordinate transformation. The answer is far 
from obvious, however geometrically we note the following relation
\begin{equation}
dx^2+dy^2 = 4 (u^2 +v^2) (du^2 +dv^2) 
\end{equation}
that essentially represents Weyl scale (or gauge) transformation. Remarks on Hopf mapping in connection with the oscillator problem
\cite{6} are also noteworthy. It seems exploring deeper geometric structure, if any, is desirable.

\section{\bf Pure gauge field and half-quantized vortex}

The significance of the circular or polar coordinate system in the emergence of the vortex structure and well-defined orbital
angular momentum states for 2D oscillator problem has been pointed out in Section 2. The duality paradigm motivates the inclusion
of polar angle in \cite{8}. An alternative new approach is proposed in the perspective of the Aharonov-Bohm effect \cite{15}.
Instead of dyon construction of \cite{8} we attribute spin-half vortex to pure gauge field.

The Schroedinger equation (14) written in polar coordinate system $(R, \phi) : 0 \leq R < \infty , 0 \leq \phi < 2 \pi$ reads
\begin{equation}
[\frac{\partial^2}{\partial R^2} + \frac{1}{R} \frac{\partial}{\partial R} +\frac{1}{R^2} \frac{\partial^2}{\partial \phi^2} 
+ \frac{2m}{\hbar^2} (E^o - \frac{m \omega^2 R^2}{2} )] \Psi^o =0
\end{equation}
Eq.(30) has the standard solution
\begin{equation}
\Psi^o_{n_r,l} (R, \phi) = C R^{|l|} e^{-\frac{m\omega}{2\hbar}R^2} ~ L^l_{n_r} (\frac{m\omega}{\hbar} R^2) e^{i l \phi} 
\end{equation}
where $C$ is a normalization constant, $L^l_{n_r}$ is associated Laguerre polynomial, $l=0,\pm 1, \pm 2,...$ and $n_r = 0, 1, 2...$. 
The eigenfunction (31) satisfies simultanoeus
eigenvalue equation for the orbital angular momentum (10). Let us consider instead $L_3 = - i \hbar \frac{\partial}{\partial \phi}$, then
\begin{equation}
L_3 \Psi^o_{n_r,l} = \hbar l \Psi^o_{n_r,l} 
\end{equation}
The corresponding energy eigenvalue is
\begin{equation}
E^o_n = \hbar \omega (|l| +1+2n_r) =\hbar \omega (n+1) 
\end{equation}
where $n= |l|+2 n_r$. The discussion in Section 2 shows that for non-zero $l$, from Eq.(4) and Eq.(7) we get the quantized vortices having 
well-defined quantized orbital angular momentum.

The correspondence with the Coulomb problem is made using the Levi-Civita transformation in polar coordinates
\begin{equation}
r = R^2,~ \theta = 2 \phi 
\end{equation}
The Schroedinger equation (30) now becomes
\begin{equation}
[-\frac{\hbar^2}{2\mu} (\frac{\partial^2}{\partial r^2} +\frac{1}{r} \frac{\partial}{\partial r} + \frac{1}{r^2}
\frac{\partial^2}{\partial \theta^2}) - \frac{\alpha}{r} ] \Psi^c = E^c \Psi^c
\end{equation}
Since $\phi \in [0, 2\pi)$ the transformed polar angle defined by (34) is $\theta \in [0, 4 \pi)$. The wavefunction becomes two-valued 
$\Psi^c(r,\theta +4\pi)=\Psi^c(r, \theta)$ and
$\Psi^c(r,\theta +2\pi)=- \Psi^c(r, \theta)$. Ter-Antonyan \cite{8} introduces inner quantum number $s=0,\frac{1}{2}$ to distinguish them, and
to account for the half-integral angular momentum. The quantum number $s$ is just an index for the wavefunction in \cite{8} such that each of its 
values represents a separate Hilbert space.

This idea has an alternative physical meaning in terms of the local gauge transformation \cite{15}
\begin{equation}
\Psi^c(r,\theta) \rightarrow e^{i s \theta} \Psi^c(r,\theta) 
\end{equation}
where the inner quantum number $s$ becomes a part of the rotation generator 
\begin{equation}
L^s_3 = - i \hbar \frac{\partial}{\partial\theta} +s 
\end{equation}
The angular momentum operator (37) has noninteger eigenvalues, and $s$ determines the different irreducible representations in the same 
Hilbert space \cite{15,16}. Note that one needs two real numbers to define rotation and translation for the group E(2) and its covering group \cite{16}.
In the present case $s=0$ defines E(2) itself while $s=\frac{1}{2}$ is 2-fold covering group of E(2). Interestingly using unitary transformation on the 
basis vectors in the Hilbert space one could construct separate Hilbert spaces for each $s$; now one has a constant phase factor $e^{i 2\pi s}$, and
the angular momentum operator takes the usual form $-i\hbar \frac{\partial}{\partial \theta}$. Two-valuedness of the wavefunction is explained by
this multiplicative phase factor for $s=0$ and $s=\frac{1}{2}$.
An important consequence of our approach is that even for $l=0$ eigenfunctions, there exists half-quantized vortex for $s=\frac{1}{2}$
corresponding to the pure gauge field similar to the one suggested for the Gordon current \cite{4}. The pure gauge field half-quantized vortex
immediately follows from the expression (4).

The local gauge transformation (36) leads to a new term in the Schroedinger equation (35), i. e. essentially the squared operator (37) divided
by $r^2$. This term is analogous to that in the Aharonov-Bohm case \cite{15} with the important difference that unlike the actual magnetic vector
potential in that case, here this term has geometric origin defining the vortex-line singularity. The vector potential is pure gauge
field in the punctured plane $R^2 -\{ 0 \}$, and quantized vortex results encircling the singularity as explained in Section 2.

Let us consider a complementary situation in view of the half-integral SU(2) symmetry group in the 2D oscillator \cite{2,6}. Let us proceed
with $F^o_1$ given by expression (10) written in circular coordinates. This generator of the rotation may formally be changed to
$-i \hbar \frac{\partial}{\partial \phi^\prime}$, where $\phi^\prime =2 \phi$. Eq.(30) can be re-expressed in terms of $\phi^\prime \in [0, 4\pi)$.
Now the local gauge transformation (36) is applied to the oscillator wavefunction
\begin{equation}
\Psi^o(R,\phi^\prime) \rightarrow e^{i s \phi^\prime} \Psi^o(R, \phi^\prime) 
\end{equation}
and the pure gauge half-quantized vortex for $l=0$ gets linked with the oscillator.

The occurrence of spin-half and the half-quantized vortex shown in oscillator and Coulomb problems constitutes a significant new result 
of the present paper. The idea that 2D oscillator symmetry group may have some connection with the electron spin \cite{1} finds realization
in the present analysis.

Another important new insight concerns the origin of charge. The origin of charge seemingly a consequence of the coordinate transformation
noted in \cite{8} is an interesting idea. Note that in \cite{5,7} dimensionless radial coordinates $\sqrt\frac{m\omega}{\hbar} R$ 
and $\frac{\hbar^2}{2\mu e^2} r$ for oscillator
and Coulomb radial equations respectively are used, and the connection between respective eigenvalues and eigenfunctions is established.
In a departure from the usual approach in \cite{8} the coupling constant is sought to be related with the energy eigenvalues in the 
defining expression (22). The energy-dependent coupling constant, and the freedom in choosing normalization length scale, let us
say, $\eta$, in (22) allow the possibility of exploring new physics. As an illustration, for $N=0$ setting $\eta$ equal to the classical
electron charge radius $\frac{e^2}{mc^2}$ and oscillator frequency $\omega = \frac{mc^2}{\hbar}$ we find 2D hydrogen atom coupling
constant
\begin{equation}
\alpha =e^2 
\end{equation}

The proposed mechanism for the origin of charge in terms of the fractional spin, see e. g. \cite{1} , can also be related with (22) in the 
followiong way. Re-writing expression (22) in the form of two angular momenta $f=\frac{e^2}{2\pi c}$ and $\hbar$ we have
\begin{equation}
 \frac{f}{\eta} = \frac{\hbar}{\lambda} 
\end{equation}
In analogy to the electron charge we introduce a charge for $\hbar$
\begin{equation}
\hbar = \frac{g^2}{2\pi c} 
\end{equation}
Note that the vortex interpretation gives natural explanation of the sign of the charge in terms of the circulation or vorticity
for vortex/anti-vortex.

An intriguing but logically justified possibility is to treat the Coulomb coupling constant $\alpha$ fixed and seek the determination
of the oscillator force constant $K=m\omega^2$
\begin{equation}
\frac{\alpha}{\eta} = \hbar \sqrt{\frac{K}{m}} 
\end{equation}
Do these speculations have physical implications? We discuss this question in the next section.

\section{\bf Discussion and Conclusion}

The role of symmetry group SU(2) for 2D isotropic oscillator has been known since the work of Jauch and Hill \cite{2}. In the
envisaged $2+1$ dimensional internal structure of the electron and neutrinos \cite{17} it was natural to anticipate the significance of
2D oscillator symmetry to explain electron spin \cite{1,3}. Analytical formulation of this idea in terms of half-quantized vortex carried out
here is a significant advancement towards this goal. The insights obtained in the Coulomb-oscillator correspondence \cite{8,9} and the 
recognition of the role of pure gauge field angular momentum in the Aharonov-Bohm effect \cite{15} have played key role in this. Note
that the 2D isotropic oscillator itself admits half-quantized vortex: consider the generator of rotation defined by (10) and make a change 
in the polar angle $\phi^\prime =2 \phi$, then the half-quantized vortex follows using the gauge transformation. The Coulomb-oscillator
correspondence throws light on the origin of the coupling constant \cite{8,9}. We have shown that the hypothesis that electron charge has origin 
in the fractional spin \cite{1} can be given a new foundation re-interpreting the relation (22).

The main motivation for the present study is to make progress in the vortex model of the electron \cite{1}, however the results
obtained here have a potential for generalization to understand elementary particle physics. An outline for the likely prospects in this 
direction is discussed below in four parts.

{\bf I}: In the duality perspective \cite{8} magnetic monopole charge having standard Dirac value is obtained. We have pointed out that
the duality paradigm has diverse manifestations \cite{18}. The proposed monopole-spin equivalence principle \cite{19} along with the 
expressions (41) and (42) could be of interest  to re-visit ideas in \cite{8,9}.

{\bf II}: The Levi-Civita coordinate transformation has proved its usefulness in the study of the 2D Coulomb-oscillator correspondence
as demonstrated in \cite{8,9}. In Section 3 we have raised a question whether the partial equivalence shows the limitation of the 
Levi-Civita transformation or it is of fundamental nature. This question deserves attention since symmetry, degeneracy and formal correspondence
contribute in these studies \cite{2,5,6,7,8,9}. In the present paper we have extended the symmetry to the gauge symmetry and obtained
a new result, namely the half-quantized vortex. Does there exist a composite coordinate and gauge transformation to answer above question?

To make physical sense we look into two important arguments. Dulock and McIntosh \cite{20} investigate degeneracy in the problem of
cyclotron motion. Authors underline the intricate role of gauge transformation, and demonstrate the translation of the center of
the orbit as a consequence of the gauge transformation. The second reason relates with the recent discussion on the orbital angular momentum
associated with the pure gauge field \cite{15,21}. In a short review \cite{21} Wakamatsu et al analyze logically
admissible arguments in this connection. In particular, their discussion on the path choice and gauge seems to indicate that enlarging
the Levi-Civita transformation with appropriate gauge transformation needs to be explored in the context of present study.

{\bf III}: Nonrelativistic Coulomb-oscillator correspondence investigated here, at first, may appear to be of no use in QCD. However on 
closer examination we find that atleast there are two aspects of QCD where the present work could be useful. First one relates with
a recent development in the non-perturbative QCD. In the basis light-front quantization (BLFQ) approach \cite{22} the crucial role of
the 2D oscillator wavefunction as the basis mode function has been highlighted. The second example  is that of the phenomenological
models to explain the hadron mass spectrum. The review on light mesons \cite{23} brings out the essentials of such models in the 
context of non-perturbative QCD. In a simplified model an approximate nonrelativistic static potential between the bound constituents
of the mesons has proved quite useful. In one of such models, a linear confinement part together with a Coulomb-like potential has
been of wide interest. Limiting cases set by QCD scale $\Lambda =200$ MeV are
\begin{equation}
V(r) = b r ~ ~ r >> \frac{1}{\Lambda} 
\end{equation}
\begin{equation}
V(r) = - \frac{4}{3} \frac{\alpha_s}{r} ~ ~ r < < \frac{1}{\Lambda} 
\end{equation}
The estimated value of the constant b is $0.18 GeV^2$, and $\alpha_s$ is the strong coupling constant. Spin-dependent interaction terms, and 
variants of this potential, for example, harmonic oscillator  potential instead of a linear confining potential have also been studied in the 
literature.

There are two attractive novel aspects of the Coulomb-oscillator correspondence relevant for the mentioned QCD problems.The energy-dependent
coupling constant, in a suitable modified form may prove useful for the QCD coupling constant. The wavefunction transformed from Coulomb to
oscillator problems may give new insights postulating internal coordinates $x_i$ and the usual space-time variables as functions of $x_i$. 
This kind of approach may have utility in BLFQ for the basis mode functions.

{\bf IV}: In the topological approach to particle physics and unified theories, the knot theory has been of great interest. However
Atiyah has drawn attention to the significant idea of Kelvin's vortex atoms in several of his papers. Though the vortex atom idea has
been discarded, the fact that stability, variety and spectrum are important characteristics of the knotted vortex tubes has been
highlighted by him, see e. g. \cite{10}. In the modern theories vortex knots have attracted a great deal of attention, however
the success has been rather very little in elementary particle physics. To make progress we suggest a revision in the outlook in which
space-time vortices are underlying fundamental structures \cite{1}. The main problem is to incorporate both global and local aspects. 
A nice way seems to be the application of Morse theory of critical points \cite{12} also discussed in \cite{10}. The present study
brings out the significance of the vortex structure in the analytical form of the wavefunctions in 2D Coulomb-oscillator correspondence.
In principle, this admits the possibility for a viable construction of knotted structures. The importance of 3-torus $T_3$ for hadron structure
suggested in \cite{19} gets support in the light of remarks made by Morse \cite{12}: $T_3$ in a sense is a product of three circles, 
and has a 0-cycle, three independent 1-cycles, three 2-cycles, and a 3-cycle having respectively the Betti numbers, 1, 3, 3, 1. Thus
on a topological 3-torus at least 8 critical points exist. The maxima, minima and saddle points define the criotical or equilibrium
points. Here we note the fact that SU(3) symmetry group arises in 3D isotropic oscillator \cite{2}, and this group has played an
important role in the hadron physics. It may be anticipated that Poincare's index theorem \cite{12}, de Rham cohomology and
non-Abelian Stokes theorem in the context of QCD \cite{11, 19, 24, 25} may provide a new framework. In this endeavour of unification
the new insights gained on the origin of charge and coupling constant in the present work and the earlier studies \cite{1} provide
a new perspective.

To conclude: half-quantized vortices and the role of pure gauge field in 2D Coulomb-oscillator correspondence give interesting new results.

\end{document}